# Claim Plane: Enforceable Change Intents and Dynamic Scope for Parallel Coding Agents

*A Deterministic Pre-Write Coordination Architecture*

**Maxim Nikolaev**
Vladivostok State University, Vladivostok, Russia
Email: m.nikolaev.contact@gmail.com | ORCID: 0009-0009-5804-8270

Code and prototype: github.com/SkeinRank/claim-plane

**Abstract.** Parallel coding agents can independently produce locally valid changes while still interfering at integration time, expanding beyond planned scope, or relying on premises invalidated by concurrent work. Existing responses emphasize communication, isolated workspaces, late merge-time repair, continuous supervision, or post-hoc runtime recovery. This paper presents Claim Plane, a model-agnostic coordination architecture that treats concurrent software change as a pre-write admission problem. Before implementation, each worker declares a versioned ChangeIntent containing an exact base commit, typed resources, dependencies, and operations marked as committed or contingent. A deterministic control plane atomically admits compatible intents, constrains same-file parallelism to declared regions, serializes unresolved overlap, tracks dependency invalidation, and fails closed on ambiguous authority. During execution, a contingent mutation does not reserve write ownership initially; the first attempted mutation triggers atomic scope promotion and re-admission against the current active set. Brokered execution binds capabilities to intent versions, leases, OS-level worktree locks, monotonic fencing tokens, and Git-tree provenance, while integration verifies immutable patches and evidence. A preliminary six-pair CooperBench mechanism check is reported only as feasibility evidence: static Claim Plane achieved 6/6 pair passes with full serialization, while dynamic scope retained parallel admission on half of the pairs, performed seven successful scope promotions, and failed closed on two undeclared mutations. The sample is intentionally too small for comparative claims. We argue that separating probabilistic planning from deterministic authority provides a foundation for a future learned semantic-dependency model and frontier-model escalation only on unresolved cases.

## 1. Introduction

Coding agents increasingly operate at repository scale rather than on isolated functions. Once several agents work on different parts of the same codebase, the central systems problem is no longer only generation quality. It is coordination under uncertainty: two workers may touch the same file for unrelated reasons, change distinct files that nevertheless share a contract, discover a dependency after implementation has started, or produce individually valid branches that fail when composed. CooperBench makes this failure mode explicit: across more than 600 collaborative coding tasks, agents show a substantial coordination penalty even when their individual task capability is stronger [1].

The emerging literature explores several complementary remedies. Centralized asynchronous delegation uses dependency-aware plans and isolated workspaces [2]; cohesion-aware orchestration partitions repositories using structural dependencies [3]; runtime substrates let supervisors inspect, fork, and replay execution traces [4]; CRDT-inspired coordination seeks deterministic convergence without explicit communication [5]; and recent concurrency-control systems notify or repair agents after shared-state interactions [6]. A particularly close line of work treats shared mutation as a pre-write admission problem [7]. These results collectively suggest that coordination should be treated as a systems layer rather than as an unstructured conversation between agents.

Claim Plane develops a specific point in this design space: enforceable, versioned change intent before and during coding. The key distinction is between what an agent is currently authorized to mutate and what it may plausibly need later. Committed scope participates in admission immediately. Contingent scope is declared in advance but does not reserve write ownership. When a worker first attempts to mutate a contingent surface, the control plane atomically promotes the requested path or bounded region and re-runs the same admission logic against the current set of active intents. The mutation proceeds only if the new authority is admitted.

This design separates two kinds of uncertainty that are often conflated. A planner or coding model may be probabilistic and imperfect when predicting future work. The authority boundary, however, need not be probabilistic. Claim Plane accepts structured declarations from a human, an LLM planner, or another tool, but admission, state transitions, scope promotion, leases, fencing, and verification are deterministic and auditable. The model may propose; the control plane decides what authority actually exists.

The paper makes four contributions:

- A ChangeIntent protocol that binds planned operations to an exact repository base and distinguishes committed authority from contingent possible scope.
- A Dynamic Scope mechanism that promotes contingent mutations just in time through atomic re-admission, preserving initial parallelism when uncertainty can be deferred.
- A deterministic enforcement path combining capability checks, intent versioning, writer leases, OS-level worktree locks, monotonic fencing tokens, broker-derived Git-tree provenance, and immutable integration evidence.
- A preliminary mechanism-oriented evaluation on six CooperBench task pairs, reported with explicit limitations, together with a research agenda for a larger frozen-plan evaluation and learned semantic dependency prediction.

The contribution is deliberately narrower than a general multi-agent orchestrator. Claim Plane does not prescribe which coding model, planner, IDE, or task decomposition strategy must be used. It is intended to sit between planning and mutation, or around existing agent runtimes, as an independent concurrency-control and evidence layer.

## 2. Problem Formulation

### 2.1 From merge conflicts to authority conflicts

Git worktrees and branches isolate bytes during generation, but they do not establish that two planned changes are compatible. A late merge conflict is only one observable symptom. More difficult cases include semantic interference without textual overlap, incompatible contract changes, stale assumptions, and scope expansion that becomes visible only after an agent has already consumed substantial inference and testing budget.

We therefore distinguish an integration conflict from an authority conflict. An integration conflict is detected after changes exist. An authority conflict exists earlier: two active workers currently claim incompatible rights or premises over a shared software surface. The purpose of pre-write admission is to decide whether that authority combination should exist before the risky mutation is accepted.

### 2.2 System model

Let $A$ be a set of coding workers operating over repository state $S$ pinned to an immutable Git base commit $b$. Each worker may be an LLM agent, a human-controlled process, or an external tool adapter. The system exposes a deterministic control plane between declared work and governed mutation. Workers may execute in isolated worktrees or sandboxes; Claim Plane does not require the model itself to run inside the control plane.

For each worker $a$, planned work is represented by a ChangeIntent $I_a$. The intent contains identity, owner, task, base revision and exact base commit, typed operations, optional dependencies, preserve policies, acceptance commands, metadata, and a liveness lease. Resources can denote files, bounded regions, symbols, concepts, contracts, routes, schemas, configuration, or documents. The current implementation is Python-first for typed callable extraction, but the protocol itself is language-agnostic.

## 3. Claim Plane Architecture

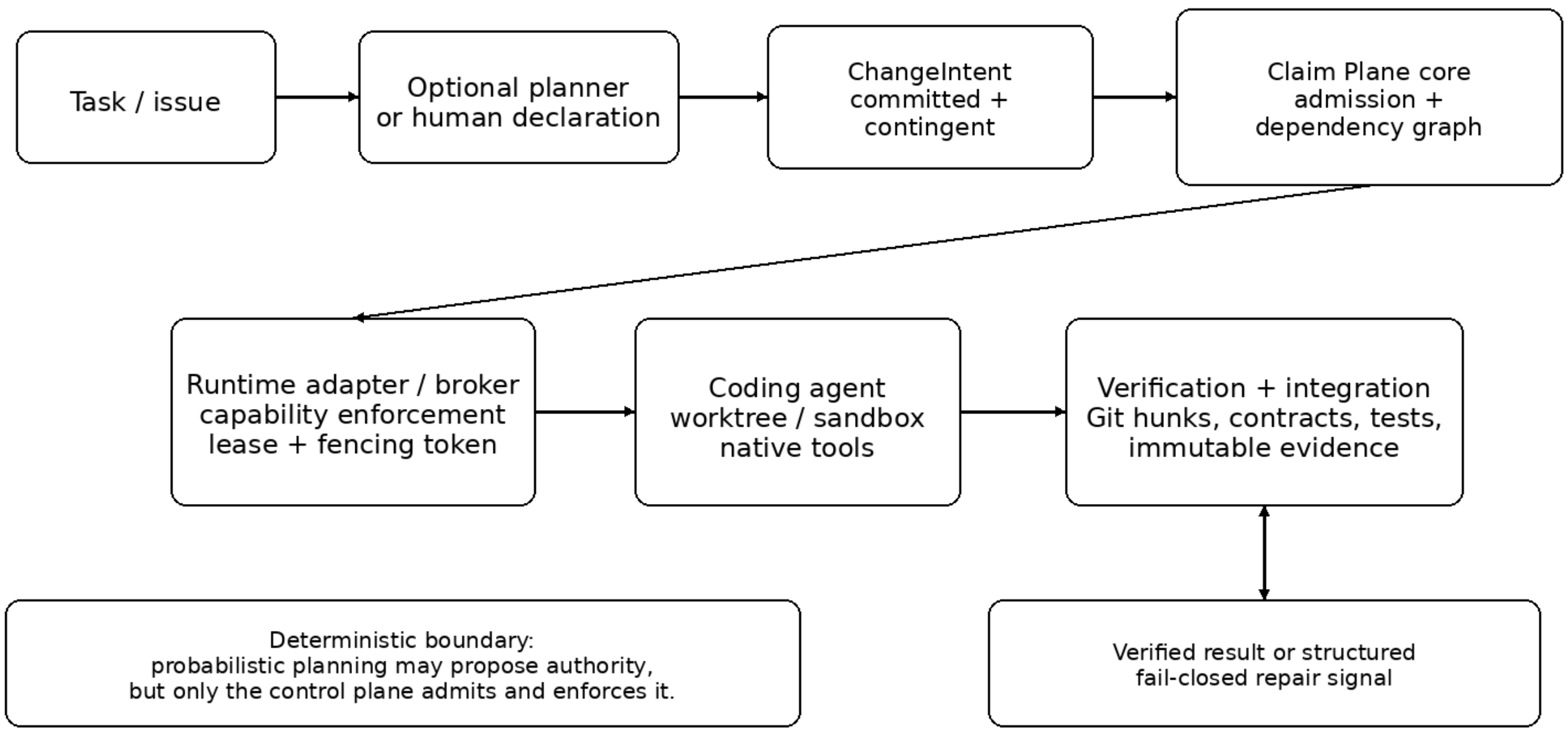

*Figure 1. Claim Plane separates probabilistic intent generation from deterministic admission, mutation authority, and verification.*

### 3.1 ChangeIntent as the unit of coordination

We model an intent as the tuple

$$I = (\text{id}, \text{task}, \text{owner}, b, O_c, O_q, D, P, T, M, l)$$

where $b$ is the exact base commit; $O_c$ is the set of committed operations; $O_q$ is the set of contingent operations; $D$ is a set of declared dependencies; $P$ is a set of preserve policies; $T$ is acceptance or test evidence requested by the caller; $M$ is metadata; and $l$ is a lease. An operation couples an access mode such as read, write, extend, delete, or rename with a typed resource reference.

The committed/contingent split is central. A committed mutation reserves current authority and participates directly in admission. A contingent mutation expresses plausible future scope. It is projected as a read premise for coordination, but it does not reserve write ownership until promotion. This allows the planner to be conservative about possible support surfaces without forcing every uncertainty into immediate serialization.

### 3.2 Atomic admission

Admission compares an incoming intent against the active intent set inside one registry transaction. The engine considers exact resource identity, broad scopes, bounded regions, semantic concepts, concept-bound contracts, destructive operations, base revisions, explicit dependencies, and graph acyclicity. Unknown overlapping writes fail closed. Broad scope overlap is not optimistically treated as independence. Same-file parallelism is allowed only when declared regions are disjoint and later verification confirms that actual Git hunks remained within those bounds.

The current decision vocabulary includes:

- independent - no relevant active overlap;
- compatible_overlap - overlap is permitted by a shared concept-bound contract;
- contract_dependency - a consumer proceeds against a producer contract and is tracked as dependent;
- parallel_with_constraint - parallel execution is allowed only within declared disjoint regions;
- notify_on_change - a premise is tracked and may invalidate a dependent;
- serialize - known conflicting writes must execute sequentially or be split;

- replan or reject - the declaration is inconsistent, ambiguous, cyclic, or otherwise unsafe.

### 3.3 Dependency invalidation

Dependencies are stored as consumer-to-producer edges while external scheduling exposes producer-first topological order. When a producer changes a resource that a consumer relied on, the first invalidation hop is filtered by the affected resource key. The consumer then becomes stale. Once stale, its own outputs are no longer trusted, so staleness propagates transitively to downstream consumers until affected work is amended and re-admitted. This avoids a global lock while still making stale premises explicit and enforceable.

## 4. Dynamic Scope: Deferred Authority with Re-Admission

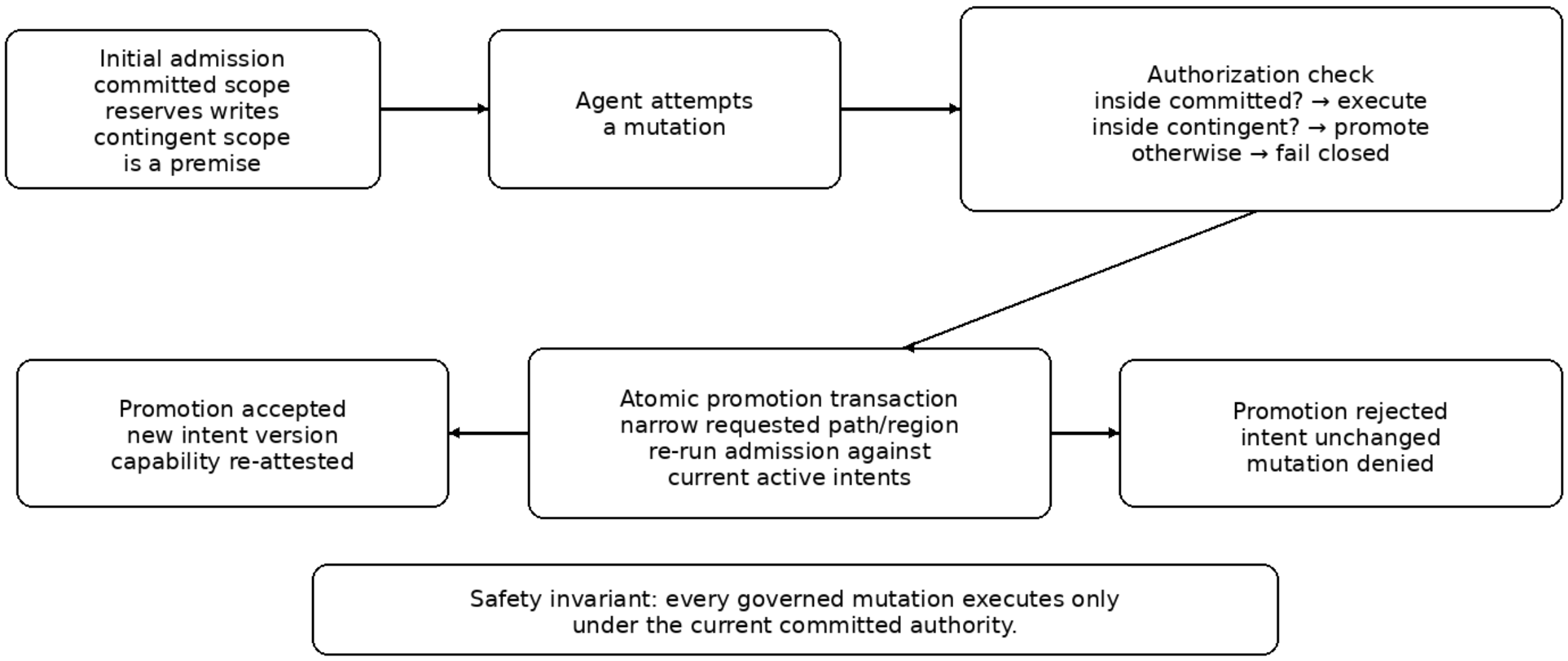

*Figure 2. A contingent mutation receives no initial write authority. The first attempted mutation must be promoted and re-admitted atomically.*

### 4.1 Promotion protocol

Dynamic Scope addresses a recurring planner trade-off. A narrow declaration improves parallelism but risks blocking a legitimate supporting edit. A broad declaration improves recall but may serialize unrelated work. Claim Plane lets a planner declare bounded uncertainty without immediately acquiring write authority.

Promotion is monotonic and atomic:

- Select a predeclared contingent operation that covers the requested path or region.
- If the contingent declaration is a broad pattern, narrow it to the concrete requested resource rather than promoting the entire pattern.
- Construct a new intent content version with the requested mutation moved into committed authority.
- Re-run admission against the current active set inside the same transaction.
- Commit the new version and re-attest the broker capability only if admission succeeds; otherwise leave the previous intent unchanged and deny the mutation.

This mechanism changes the timing of uncertainty. The planner no longer has to choose between immediate pessimism and unbounded optimism. It can surface potential support regions as contingent and let the runtime reveal which uncertainty actually becomes operational.

### 4.2 Region-aware authorization and coordinate stability

Line-region authority introduces a practical problem: after earlier edits, current working-tree line numbers may no longer match the base image used by the planner. Claim Plane therefore treats the planner/base image as the authorization coordinate system. A runtime adapter maps current mutation coordinates conservatively back to pre-image coordinates before checking committed or contingent authority. Equal blocks map line-for-line; inserted blocks map to a nearby pre-image anchor; unequal replacements map conservatively to the replaced pre-image span. The mapping is an adapter concern rather than a weakening of the admission rule.

The underlying safety rule remains containment: an actual mutation must be contained by current committed authority after any accepted promotion. Widening a planned range merely because the agent missed by several lines would silently change the authorization policy and is therefore not used as a runtime repair.

# 5. Deterministic Enforcement and Evidence

## 5.1 Brokered execution

A control plane can only enforce pre-write authority when the governed mutation crosses an observable boundary. Claim Plane therefore supports a brokered execution mode in which file operations are performed by a trusted proxy rather than directly by the worker. The broker exposes exact capabilities: ordinary write does not imply delete or rename, append-only extension is distinct from replacement, and rename destinations must be declared.

Every request revalidates the current intent state, content version, exact base commit, repository identity, lease, and capability. A durable prepare record is written before the filesystem effect. The committed operation is then bound to a trusted observation-event range and to the broker instance that performed it. This ordering prevents a generic permission or hand-authored trace from being presented later as proof of an authorized mutation.

## 5.2 Writer authority: lease, OS lock, and fencing

For a single physical worktree, Claim Plane layers three protections. First, the registry grants one active writer lease. Second, an OS-level lock under Git's canonical common directory prevents two local control-plane processes from governing the same worktree through separate SQLite databases or linked worktree paths. Third, every new writer lease receives a larger monotonic fencing token. The token is attached to broker instances, operations, observations, and verification evidence, so a superseded writer fails closed even if it remains alive.

## 5.3 Git-tree provenance and immutable integration

Brokered mutations advance a compare-and-swap Git-tree chain. Before each operation, the broker verifies the expected pre-tree; after the mutation it checks the changed path and records the new tree identity. Any out-of-band tracked or non-ignored mutation breaks the chain and causes subsequent requests or integration to fail.

Integration is performed from immutable worker snapshots rather than mutable worktrees. For each worker, the runner freezes an exact Git tree, persists one binary patch and digest, collects a manifest from that same snapshot, runs acceptance checks without allowing the snapshot to mutate, verifies all manifests together, and applies the same persisted patch bytes in producer-first order. The resulting evidence binds the specification, base tree, worker patches, manifests, reports, result tree, and result commit. This closes a common time-of-check/time-of-use gap between verification and integration.

# 6. Formal Safety Properties

## 6.1 Admission predicate

Let $\Gamma$ be the set of active admitted intents. For an incoming intent $I$, define $\mathrm{Conflict}(I, J)$ as a deterministic predicate over current committed write authority, broad-scope overlap, bounded-region intersection, destructive operations, contract compatibility, base identity, and explicit dependency obligations. Let $\mathrm{DAG}(\Gamma \cup \{I\})$ denote acyclicity of the candidate dependency graph. Initial admission is allowed only when every unresolved conflict is routed to an accepted constrained mode and the graph remains acyclic.

Let $\mathrm{SafePair}(I, J)$ hold when the pair has no unresolved incompatible authority conflict after applying any admitted dependency or constrained-parallel relation.

$$\mathrm{Admit}(I \mid \Gamma) = \mathrm{allowed} \Leftrightarrow (\forall J \in \Gamma, \mathrm{SafePair}(I, J)) \wedge \mathrm{DAG}(\Gamma \cup \{I\})$$

The exact SafePair outcome may be independent, compatible overlap, contract dependency, or constrained parallelism. A known incompatible overlap instead yields serialization, replanning, or rejection.

### 6.2 Promotion invariant

For a mutation $m$ attempted by intent $I$, let Covers$(O_c, m)$ mean that the current committed operation set authorizes $m$. Let Promotable$(O_q, m)$ mean that a contingent operation covers $m$. The runtime rule is:

$$\text{execute}(m) \text{ only if}$$
$$\text{Covers}(O_c, m) \lor [\text{Promotable}(O_q, m) \land \text{Admit}(\text{Promote}(I, m) \mid \Gamma) \text{ succeeds atomically}]$$

A rejected promotion leaves $I$ unchanged. Therefore failed speculation cannot silently expand existing authority. Under a complete broker boundary, every executed mutation is attributable to the latest admitted intent version and current fencing token.

### 6.3 Informal safety proposition

Proposition. Assume (i) all governed repository mutations pass through the broker, (ii) the registry transaction is atomic, (iii) the OS worktree lock and fencing token are enforced, and (iv) the verifier checks the final broker-derived tree. Then no governed mutation can be accepted outside the committed authority of the current admitted intent version.

Proof sketch. A direct committed mutation passes only after capability revalidation against the current intent version and fencing token. A contingent mutation cannot execute until an atomic promotion transaction creates a newly admitted content version and re-attests the broker. An undeclared mutation is denied. A stale or superseded broker fails capability or fencing validation. Finally, an out-of-band mutation breaks the expected Git-tree chain and is rejected during the next operation or integration. The proposition does not cover bypass channels outside the declared broker or OS monitor boundary.

## 7. Prototype Implementation

The current open-source prototype is implemented in Python and exposes a CLI, stdio MCP interface, JSON Schemas, SQLite reference storage, and a model-agnostic integration runner. The base package has no required model runtime. Optional semantic resolution can consume a published Agent Lexicon vocabulary, while planners and repair agents remain external.

The public implementation supports exact files, globs, bounded regions, symbols, concepts, concept-bound contracts, routes, schemas, configuration, and documents; versioned intent amendments; dependency invalidation; trusted observation sessions; brokered reads and writes; writer leases, OS locks, and fencing; immutable acceptance execution; and SHA-256/HMAC or optional public-key evidence attestation. The strict broker boundary is Linux-first because non-bypassable filesystem isolation relies on Linux sandbox primitives, while the broker and verification pipeline can operate on macOS with weaker process-isolation assumptions.

This implementation boundary matters for product integration. A native coding runtime such as Codex, Claude Code, or an internal agent can use Claim Plane through an adapter. When a runtime exposes tool-call hooks, the adapter can enforce pre-write capabilities directly. When it does not, the system can still use isolated worktrees, pre-task admission, continuous diff verification, and merge gating, but the guarantee is correspondingly weaker because direct host writes may bypass the control plane.

## 8. Preliminary Mechanism Check

### 8.1 Purpose and setup

The evaluation in this paper is intentionally preliminary. Its purpose is to verify that the implemented mechanisms activate on live coding-agent trajectories, not to establish a statistically reliable advantage over competing systems. We use six fixed development pairs from CooperBench: three labeled conflict pairs and three labeled clean pairs. All individual gold features pass the benchmark's native checks before LLM execution.

Four arms are compared: naive parallel execution; Claim Plane with all final planned scope treated as committed (static); Claim Plane preserving committed versus contingent scope (dynamic); and always-serial execution. A single coder seed is used. The planner is DeepSeek V4 Pro and the coding agent is DeepSeek V4 Flash. The static and dynamic arms share exactly the same calibrated planner output for each pair. API calls are physically sequential, so the experiment does not measure real wall-clock parallel speedup. Initial agent context is localized using benchmark gold-patch locations, which must be disclosed as an oracle-assisted context condition.

| Arm | Pair pass | Integration | Initial serialization | Promotions | Undeclared blocks |
|---|---|---|---|---|---|
| Parallel | 2/6 (33.3%) | 3/6 (50.0%) | 0/6 (0%) | 0 | 0 |
| Claim Plane static | 6/6 (100%) | 6/6 (100%) | 6/6 (100%) | 0 | 0 |
| Claim Plane dynamic | 3/6 (50.0%) | 4/6 (66.7%) | 3/6 (50%) | 7 | 2 |
| Always serial | 4/6 (66.7%) | 6/6 (100%) | 6/6 (100%) | 0 | 0 |

*Table 1. Preliminary six-pair mechanism check. These development-set numbers are not a publication-grade effect estimate.*

### 8.2 Observations

The static arm demonstrates the safety end of the design trade-off: it achieved a clean pair pass on all six development pairs, but only by serializing all six. Dynamic scope reduced initial serialization to three of six pairs and produced seven successful just-in-time promotions without rejected promotions. Two trajectories were stopped by fail-closed undeclared-scope blocks. The dynamic arm therefore activated the intended mechanism, but it did not yet recover the static arm's quality.

Among the four pairs for which both independent parallel agents passed their own feature tests, the dynamic gate had no observed false negatives with respect to the live parallel outcome in this single run, but the eligible sample is only four pairs. One safe parallel case was conservatively serialized. This should be read as a debugging signal, not as an estimate of gate accuracy.

The planner calibration also became deliberately high-recall: it added many bounded contingent regions. This reduces undeclared writes but can make static policy extremely conservative and can reduce region precision. The larger follow-up study therefore freezes the planner outputs once and varies only the coding-agent seeds, so planner variance is not confounded with coordination policy.

### 8.3 Planned confirmatory evaluation

The next study is pre-specified as 30 frozen CooperBench feature pairs, balanced between benchmark conflict and clean labels, with one frozen Planner v1 output per feature and three independent coding-agent seeds. The four execution arms are retained. Results will be checkpointed in shards and aggregated with feature-pair and stricter task-cluster bootstrap intervals because multiple feature pairs may originate from the same underlying CooperBench task. No planner-policy tuning will be performed after observing the confirmatory outcomes except for clearly identified harness bugs. This larger experiment is intentionally reserved for a subsequent empirical paper rather than used to overstate the present design paper.

## 9. Related Work and Positioning

### 9.1 Benchmarks and orchestration

CooperBench [1] directly motivates the coordination problem studied here by showing that agents lose substantial capability when working together. CAID [2] uses a central manager, dependency-aware planning, isolated workspaces, and structured integration. Co-Coder [3] frames task decomposition as a graph-partitioning problem and uses static dependencies to balance parallelism against communication. Claim Plane is complementary: it assumes that tasks or intents already exist and asks whether their declared and evolving mutation authority may coexist.

### 9.2 Runtime supervision, repair, and convergence

Shepherd [4] provides a formalized execution trace that enables meta-agents to fork and replay target-agent state; a live supervisor substantially improves CooperBench pair success in its evaluation. Claim Plane instead seeks to make the common admission path deterministic and reserve expensive model reasoning for uncertain cases. CodeCRDT [5] coordinates through observable shared state and CRDT convergence. CoAgent [6] takes an optimistic runtime approach: it fixes a serialization order, notifies affected agents, and relies on agent judgment plus compensating operations to repair stale interactions. Claim Plane moves the primary enforcement boundary earlier, before a governed mutation receives authority, while still supporting dependency invalidation and re-admission during execution.

### 9.3 Closest pre-write and control-plane work

ATM [7] is the closest contemporaneous work. It explicitly formulates governed shared mutation as a pre-write admission problem and uses semantic atoms, bounded regions, a CID broker, fail-closed routing, and a neutral steward. Claim Plane does not claim priority over the general idea of pre-write admission. Its distinctive formulation is a versioned ChangeIntent with committed versus contingent authority, atomic just-in-time promotion and re-admission, intent-bound broker capability re-attestation, explicit writer leases and fencing, and an integration path that verifies the exact immutable worker patches. The two systems should be viewed as independent points in a rapidly emerging design space rather than as identical implementations.

Madatha [8] also uses the phrase deterministic control plane for LLM coding agents, but studies a different layer: agent configuration supply chains, permission boundaries, phase state machines, and prompt/configuration drift. Claim Plane focuses on concurrent software mutation authority and integration evidence. This distinction is why the present paper avoids claiming a generic "deterministic control plane" as a unique label.

At the agent-runtime level, SWE-agent [9] shows that the agent-computer interface itself materially affects software engineering performance. Claim Plane follows the same systems intuition: reliability depends not only on the model, but also on the execution interface and the invariants enforced around tool actions.

| Approach | Primary intervention point | Shared-state mechanism | LLM in control loop? | Claim Plane distinction |
|---|---|---|---|---|
| CAID [2] | Task decomposition / integration | Isolated workspaces + manager | Yes | Admission of evolving mutation authority |
| Co-Coder [3] | Partition and schedule | Dependency graph | Yes | Runtime enforcement after intent declaration |
| Shepherd [4] | Execution trace / supervision | Forkable typed trace | Yes, supervisor | Deterministic common path before mutation |
| CodeCRDT [5] | Concurrent shared state | CRDT convergence | Not necessarily | Semantic/resource authority rather than text convergence |
| CoAgent [6] | Runtime interaction / repair | Notifications + compensating actions | Yes, repair judgment | Fail-closed pre-write admission |
| ATM [7] | Pre-write admission | CID/atom broker + steward | Optional | Committed/contingent Dynamic Scope + promotion |

*Table 2. Positioning relative to representative recent systems. The categories are simplified and not mutually exclusive.*

## 10. Research Agenda: From Conservative Planning to Learned Semantic Dependency

The current reference planner is intentionally conservative because a missed dependency can permit unsafe parallel work. The natural next step is not to place more authority inside the planner, but to add a learned semantic-dependency model (SDM) as a sensor between deterministic filtering and expensive frontier supervision.

A target three-level architecture is:

```
Level 1: deterministic Claim Plane rules handle provable independence, overlap, contracts,
declared dependencies, and scope.
Level 2: a compact SDM classifies structurally suspicious candidate pairs as independent,
A-before-B, B-before-A, hard conflict, commutative, or uncertain, and emits structured
```

```
evidence with confidence.
Level 3: a frontier supervisor receives only low-confidence cases together with Claim
Plane evidence and proposes a structured resolution that still requires deterministic
admission.
```

The model should predict structure rather than hold final authority. A useful normalized representation is language-agnostic at the control-plane level: symbols, contracts, schemas, concepts, provides/requires relationships, modifies/uses relationships, and invalidation edges. Language adapters can extract these signals from Python, TypeScript, Rust, Java, Swift, or C++, while a shared backbone can learn cross-language dependency patterns. Language-specific adapters or heads may recover specialist accuracy without losing mixed-stack reasoning.

Claim Plane also creates a potential data flywheel. Each run already yields task descriptions, planner intents, declared scope, actual writes, promotions, blocks, integration outcomes, tests, and serial-rescue outcomes. These traces can be converted into labels such as parallel succeeded, serialization rescued, dependency direction worked, or scope promotion failed. The highest-cost part of an SDM project is likely to be trustworthy labels rather than raw GPU time; a control plane that observes real execution can generate those labels as a by-product of use.

## 11. Limitations and Threats to Validity

First, Claim Plane cannot enforce mutations that bypass its observable boundary. Broker-complete or OS-monitored execution supports stronger claims than post-hoc diff verification. Product integrations must therefore state their enforcement tier explicitly.

Second, the current prototype is Python-first for typed symbol and callable extraction. The protocol is language-agnostic, but high-quality semantic analysis requires language adapters and, eventually, multilingual training data.

Third, the preliminary experiment uses only six development pairs, one coder seed, oracle-localized initial context, and logically parallel but physically sequential API execution. The reported percentages are not statistical evidence and cannot be compared directly with published results from different task sets, models, or harnesses.

Fourth, planner coverage and planner precision remain a central trade-off. More contingent regions reduce undeclared writes but can create noisy authority candidates and increase conservatism. Dynamic scope defers that cost but does not eliminate the need for better semantic dependency prediction.

Fifth, the reference SQLite lease and OS lock provide single-host authority. Multi-host deployment requires a network-authoritative store and distributed lease/fencing sequence while retaining local host locks.

Finally, pre-write admission is not a substitute for Git merge, tests, code review, or behavioral verification. It is one additional enforcement layer intended to prevent or surface unsafe concurrency earlier.

## 12. Conclusion

Claim Plane reframes parallel coding-agent coordination as a problem of enforceable change authority. A probabilistic planner may predict what a worker is likely to touch, but deterministic admission decides what the worker is authorized to mutate. Committed scope provides immediate authority; contingent scope preserves uncertainty without reserving writes; Dynamic Scope promotes only the resource that becomes necessary and re-runs admission before the mutation proceeds. Leases, OS locks, fencing tokens, broker provenance, and immutable verification connect that abstract intent to real repository state.

The preliminary CooperBench mechanism check shows that the architecture is executable and that runtime promotion occurs in live agent trajectories, while also exposing the present planner's conservatism and remaining scope misses. The next empirical stage is a frozen-plan 30-pair, three-seed evaluation. Longer term, the architecture suggests a layered path in which deterministic rules handle what can be proven, a learned semantic-dependency model handles what can be predicted with confidence, and frontier intelligence is reserved for the unresolved edge cases.

## Reproducibility and Tool-Use Statement

The Claim Plane prototype and protocol documentation are publicly available at the repository listed on the first page. The preliminary benchmark artifacts used for the mechanism check include selected task pairs, configuration, planner declarations, agent traces, dynamic-scope events, and aggregate metrics. Generative AI tools were used as assistive tools during software development and manuscript preparation. All technical claims, references, experimental results, and final text were reviewed and validated by the author.

## Appendix A. Reference Admission and Promotion Pseudocode

```
function ADMIT(intent I, active intents Γ):
    canonicalize I
    verify exact base commit and referenced dependencies
    conflicts ← compare committed authority of I against Γ
    graph ← candidate dependency graph Γ ∪ {I}
    if graph contains cycle:
        return REJECT
    if unresolved incompatible overlap exists:
        return SERIALIZE or REPLAN
    persist admitted intent version atomically
    return ALLOW(decision, constraints, notices)

function AUTHORIZE_MUTATION(intent I, mutation m):
    if committed_scope(I) covers m:
        validate lease, intent version, capability, fencing token
        execute through broker and record tree transition
        return SUCCESS
    if contingent_scope(I) covers m:
        I' ← promote only the concrete path/region required by m
        if ADMIT(I', current active intents) succeeds atomically:
            re-attest broker to I'
            execute m
            return SUCCESS
        return BLOCK
    return BLOCK
```